\documentclass[aps,prd,12pt,notitlepage,showpacs]{revtex4-1}
\usepackage{graphicx}
\usepackage{amsmath}
\usepackage{amssymb}
\usepackage{mathrsfs}
\usepackage{amsfonts}
\usepackage{dsfont}
\usepackage{dcolumn}
\usepackage{bm}
\usepackage{booktabs}
\usepackage{textcomp}
\usepackage{color}
\usepackage{xcolor}
\usepackage[retainorgcmds]{IEEEtrantools}
\usepackage{bigints}

\providecommand{\abs}[1]{\left\lvert#1\right\rvert}
\providecommand{\pd}[2]{\frac{\partial#1}{\partial#2}} 
\providecommand{\fracd}[2]{\frac{\displaystyle{#1}}{\displaystyle{#2}}}

\providecommand{\ket}[1]{|#1\rangle}
\providecommand{\bra}[1]{\langle#1|}
\providecommand{\brak}[2]{\langle#1|#2\rangle} 

\newcommand{\mL}{\mathcal{L}}

\newcommand{\bep}{\bm{\epsilon}}
\newcommand{\br}{\mathbf{r}}
\newcommand{\vx}{\mathbf{x}}
\newcommand{\vy}{\mathbf{y}}
\newcommand{\vp}{\mathbf{p}}
\newcommand{\vq}{\mathbf{q}}

\newcommand{\vP}{\mathbf{P}}
\newcommand{\va}{\mathbf{a}}
\newcommand{\vn}{\mathbf{n}}

\newcommand{\bphi}{{\phi^*}}
\newcommand{\bnabla}{\bm{\nabla}}

\newcommand{\hphi}{\hat{\phi}}

\newcommand{\hbphi}{{\hat{\phi}^\dagger}}
\newcommand{\hH}{\hat{H}}
\newcommand{\hP}{\hat{P}}
\newcommand{\hN}{\hat{N}}

\newcommand{\ha}{\hat{a}}
\newcommand{\hac}{{\hat{a}^\dagger}}

\newcommand{\hPhi}{\hat{\Phi}}
\newcommand{\hbPhi}{{\hat{\Phi}^\dagger}}
\newcommand{\npa}{{| \mathbf{p}_1, \ldots, \mathbf{p}_n\rangle}}

\newcommand{\nXt}{{| \mathbf{x}_1, \ldots, \mathbf{x}_n; z\rangle}}

\usepackage[all]{xy}

\begin{document}

\setlength{\baselineskip}{.85\baselineskip}
%
%

\title{\large Field theory of monochromatic  optical beams. II \\ {\textsl{Classical and quantum paraxial fields}}}
\author{Andrea Aiello}\email{andrea.aiello@mpl.mpg.de}
\affiliation{Max Planck Institute for the Science of Light,\\
G\"{u}nter-Scharowsky-Stra{\ss}e 1/Bau 24, 91058 Erlangen, Germany}
\affiliation{Institute for Optics, Information and Photonics, \\
 University of Erlangen-Nuernberg, Staudtstrasse 7/B2, 91058 Erlangen, Germany}

\begin{abstract}
This work is the second part of an investigation aiming at the study of optical wave equations from a field-theoretic point of view. Here, we study classical and quantum aspects of scalar fields satisfying the paraxial wave equation. First, we determine conservation laws for energy, linear and angular momentum of paraxial fields in a classical context. Then, we proceed with the quantization of the field. Finally, we compare our result with the traditional ones.  
\end{abstract}
\pacs{xx.xx.Aa}

\date{\today}
\maketitle

\section{Introduction}

In this second paper of the series ``Field theory of monochromatic  optical beams,'' we continue the investigation of scalar fields obeying either the Helmholtz wave equation (HWE)
\begin{align}
\left( \frac{\partial^2}{ \partial x^2} + \frac{\partial^2}{ \partial y^2} + \frac{\partial^2}{ \partial z^2} + k^2_0 \right)\psi(\vx,z) = 0, \qquad k_0>0,\label{f10}
\end{align}
and the paraxial wave equation (PWE)
\begin{align}
\left( \frac{\partial^2}{ \partial x^2} + \frac{\partial^2}{ \partial y^2} + 2 i k_0 \frac{\partial}{ \partial z}  \right)\phi(\vx,z) = 0, \qquad k_0>0,\label{f20}
\end{align}
with $\vx = (x,y) \in \mathbb{R}^2$. Specifically, this work is devoted to the study of some properties of paraxial fields, in both classical and quantum regimes. 

The notation that we use here, is the same as established in part I. The three-dimensional gradient is expressed as $\partial \psi/ \partial x^\mu \equiv \partial_\mu \psi = \left(\bm{\nabla},\partial_z \right) \psi$, where a point in $\mathbb{R}^3$ is labeled by the three coordinates $x^\mu$, with $x^3 = z$ the \emph{longitudinal} coordinate and $x^k, \; k=1,2$ the \emph{transverse} coordinates.
The \emph{two-dimensional} gradient of a scalar function $f(x,y,z)$ is denoted $\bm{\nabla} f$ and is defined as
\begin{align}
\bm{\nabla} f = \frac{\partial f}{\partial x} \bep_1 + \frac{\partial f}{\partial y} \bep_2,\label{f65}
\end{align}
where   $\bep_1$ and $\bep_2$ are the orthogonal unit vectors pointing in the $x$ and $y$ Cartesian coordinate directions, respectively.
Greek indexes $\mu, \nu,\alpha,\beta, \ldots$ , run from $1$ to $3$, while Latin indexes $i,j,k,l,m,n, \ldots$ , take the values $1$ and $2$. Moreover, $\partial^2 = \partial_x^2 + \partial_y^2 + \partial_z^2$ and  $\nabla^2 = \partial_x^2 + \partial_y^2$.

\section{Two Lagrangians for a paraxial field}

Let $\phi(\vx,z)$ be a complex scalar field satisfying the paraxial wave equation, namely
\begin{align}\label{p10}
\left( i \pd{}{z} + \frac{1}{2 k_0}\nabla^2  \right)\phi(\vx,z) = 0,
%
%
\end{align}
which is reminiscent of the Schr\"{o}dinger equation for a free particle on a plane. A suitable Lagrangian generating Eq. \eqref{p10} should be bilinear in the field and its derivatives:
\begin{align}\label{p20}
\mL = A \, \bphi \partial_3 \phi + B \, \phi \partial_3 \bphi + C \bphi \phi + D \, \delta^{ij} \partial_i \bphi \partial_j \phi,
%
%
\end{align}
where the four coefficients $A,B,C,D$ are determined by imposing the fulfillment of the  Euler-Lagrange equation
\begin{align}\label{p30}
\frac{\partial \mL}{\partial \bphi} - \frac{\partial}{\partial x^\mu} 
\frac{\partial \mL}{\partial (\partial_\mu \bphi)} =  0.
%
%
\end{align}
A straightforward calculation shows that substituting Eq. \eqref{p20} into Eq. \eqref{p30}, one obtains
\begin{align}\label{p40}
\left( A - B \right) \partial_3 \phi - D \, \nabla^2 \phi +  C  \phi =0.
%
%
\end{align}
Now, requiring the equality between Eq. \eqref{p10} and Eq. \eqref{p40} yields the following relations:
\begin{align}\label{p50}
 A - B = i, \qquad D = - \frac{1}{2 k_0}, \qquad C=0.
%
%
\end{align}
The equation $ A - B = i$ can be satisfied with different choices of $A$ and $B$. 
 We distinguish between the \emph{symmetric} choice $A = - B = i/2$, leading to the Lagrangian $\mL_1$, and the asymmetric choice $A = i, \, B=0$, which generates the Lagrangian $\mL_2$, where 
\begin{align}\label{p60}
\mL_1 = \frac{i}{2} \bigl(  \bphi \partial_3 \phi - \phi \partial_3 \bphi \bigr)  - \frac{1}{2 k_0} \delta^{ij} \partial_i \bphi \partial_j \phi
%
%
\end{align}
and 
\begin{align}\label{p70}
\mL_2 = i \bphi \partial_3 \phi  - \frac{1}{2 k_0} \delta^{ij} \partial_i \bphi \partial_j \phi.
%
%
\end{align}
The first Lagrangian $\mL_1$ is much more appealing and it is clearly real, while $\mL_2$ is not. However, $\mL_1$ and $\mL_2$ differ by a total $z$-derivative which does no affect the dynamics:
\begin{align}\label{p80}
\mL_2 - \mL_1 = \frac{i}{2} \partial_3 \bigl( \bphi \phi \bigr).
%
%
\end{align}
In fact, as we shall see soon, only $\mL_2$ leads to the correct equations in the Hamilton form.

\subsection{First Lagrangian: $\mL_1$}

In this case there are two \emph{independent} fields $\Pi$ and $\Pi^*$ canonically conjugate to $\phi$ and $\bphi$, respectively, specifically

\begin{align}\label{p90}
\Pi_1 = \pd{\mL_1}{(\partial_3 \phi)} = \frac{i}{2} \phi^*, \qquad \Pi^*_1 = \pd{\mL_1}{(\partial_3 \psi^*)} = -\frac{i}{2} \phi.
%
%
\end{align}
The \emph{Hamiltonian density} $\mathscr{H}_1$ is defined  in the standard way: 
\begin{align}\label{p100}
\mathscr{H}_1 = & \; \Pi_1 \, \partial_3 \phi + \Pi_1^* \partial_3 \phi^* - \mL_1 \nonumber \\
= & \; \frac{1}{2 k_0} \bnabla \phi^* \cdot \bnabla \phi \nonumber \\
= & \; -\frac{i}{ k_0} \, \delta^{ij} \partial_i \Pi_1 \partial_j \phi .
%
%
\end{align}
The \emph{total} Hamiltonian $H_1$ is simply 
\begin{align}\label{p110}
H_1 = & \; \int d \vx \, \mathscr{H}_1 \nonumber \\
= & \; -\frac{i}{ k_0} \, \delta^{ij}\int d \vx \,  \partial_i \Pi_1 \partial_j \phi\nonumber \\
= & \; \frac{i}{ k_0} \, \int d \vx \,   \Pi_1 \nabla^2  \phi ,
%
%
\end{align}
where to obtain the last line, integration by part has been used and a surface term has been discarded. Then, the Hamilton equations give
\begin{align}\label{p120}
\pd{}{z}\phi(\vx,z) = & \; \frac{\delta H_1}{\delta \Pi_1(\vx,z)}\nonumber \\
= & \; \frac{i}{ k_0} \nabla^2  \phi.
%
%
\end{align}
It is clear that Eq. \eqref{p120} does not reproduce correctly Eq. \eqref{p10} and, therefore, $\mL_1$ must be ruled out.

\subsection{Second Lagrangian: $\mL_2$}

In this case we have 

\begin{align}\label{p130}
\Pi_2 = \pd{\mL_2}{(\partial_3 \phi)} = i \phi^*, \qquad \Pi^*_2 = 0.
%
%
\end{align}
As explained in \cite{Greiner,Schiff}, since $\Pi_2(\vx,z)$ is simply proportional to the conjugate of  $\phi(\vx,z)$, then there are only two independent fields, namely $\phi(\vx,z)$ and $\Pi_2(\vx,z)$. Therefore, the Hamiltonian density is calculated as
\begin{align}\label{p140}
\mathscr{H}_2 = & \; \Pi_2 \, \partial_3 \phi  - \mL_2 \nonumber \\
= & \; \frac{1}{2 k_0} \bnabla \phi^* \cdot \bnabla \phi \nonumber \\
= & \; -\frac{i}{ 2k_0} \, \delta^{ij} \partial_i \Pi_2 \partial_j \phi .
%
%
\end{align}
It should be noticed that the second line of Eq. \eqref{p140} coincides with the second line of Eq. \eqref{p100}.
A straightforward calculation shows that using $H_2$, the Hamilton equations 
give the correct equations of motion:
\begin{align}\label{p150}
\pd{}{z}\phi(\vx,z) = & \; \frac{\delta H_2}{\delta \Pi_2(\vx,z)}\nonumber \\
= & \; \frac{i}{ 2k_0} \nabla^2  \phi,
%
%
\end{align}
where 
\begin{align}\label{p160}
H_2 =  \int d \vx \, \mathscr{H}_2. 
%
%
\end{align}
Therefore, from now on we will consider only $\mL_2$ as the ``true'' Lagrangian for the PWE and we will drop the subscript ``$2$'' writing simply $\mL$ instead of $\mL_2$. 

The asymmetry of $\mL$ with respect to the Cartesian coordinates $x,y,z$, can be made more manifest by rewriting Eq. \eqref{p70} as
\begin{align}\label{p165}
\mL = & \; i \bphi \partial_3 \phi  - \frac{1}{2 k_0} \delta^{ij} \partial_i \bphi \partial_j \phi \nonumber \\
 = & \; i\delta^{3 \mu} \bphi \partial_\mu \phi  - \frac{1}{2 k_0} \bigl[ \delta^{ij} \partial_i \bphi \partial_j \phi + 
\left( \partial_3 \bphi \partial_3 \phi -  \partial_3 \bphi \partial_3 \phi  \right) \bigr] \nonumber \\
 = & \; i\delta^{3 \mu} \bphi \partial_\mu \phi  - \frac{1}{2 k_0} \bigl( \delta^{\mu \nu} - \delta^{3 \mu} \delta^{3 \nu}\bigr) \partial_\mu \bphi \partial_\nu \phi \nonumber \\
\equiv & \; i\delta^{3 \mu} \bphi \partial_\mu \phi  - \frac{1}{2 k_0}  \delta^{\mu \nu}_\text{T} \partial_\mu \bphi \partial_\nu \phi,
%
%
\end{align}
where $\delta^{\mu \nu}_\text{T} \equiv \delta^{\mu \nu} - \delta^{3 \mu} \delta^{3 \nu}$ is a \emph{transverse Kronecker delta}, which can also be seen as the coordinate-component of the  dyadic $\bep_1 \bep_1 + \bep_2 \bep_2$, namely $\delta^{\mu \nu}_\text{T} = \left( \bep_1 \bep_1 + \bep_2 \bep_2 \right)^{\mu \nu}$. By definition, $\delta^{3 \nu}_\text{T} = 0 = \delta^ {\mu 3}_\text{T}$, therefore $\delta^{\mu \nu}_\text{T} \partial_\mu \bphi \partial_\nu \phi = \delta^{ij} \partial_i \bphi \partial_j \phi$.

\section{Symmetries and conservation laws}

The Helmholtz equation does not contain explicitly the three Cartesian coordinates $x,y,z$. Moreover, the latter  enter in a symmetric manner in the differential operator $\partial^2 = \partial^2_x + \partial^2_y + \partial^2_z$. This yields to the invariance of the HWE under \emph{translations} and \emph{rotations} of the fields \cite{Sudbery}. Conversely, due to its first-order form in the $z$-coordinate, we do not expect to keep rotational invariance around an \emph{arbitrary} axis for the paraxial wave equation. 
In order to illustrate the symmetries exhibited by the PWE, let us consider the field $\phi(\vx,z)$ evaluated in the generic point $\mathbf{r}=(\vx,z)$ and imagine to perform an \emph{active transformation} that converts, via a translation by $\mathbf{a} = a^\mu \bep_\mu$ and a three-dimensional rotation by $\Lambda^{\mu}_{\phantom{x}\nu}$, the original field $\phi(\vx,z)$ into the new field $\phi'(\vx,z)$:
\begin{align}\label{p170}
\phi(\vx,z) \to \phi'(\vx,z).
%
%
\end{align}
Let $\mathbf{r}'=(\vx',z')$ be the point obtained by translating and rotating the original point $\mathbf{r}=(\vx,z)$ by $\mathbf{a}$ and $\Lambda$, respectively, that is:
\begin{align}\label{p180}
{x'}^\mu = \Lambda^{\mu}_{\phantom{x}\nu} x^\nu + a^\mu \qquad \Leftrightarrow \qquad \mathbf{r}' = \Lambda \mathbf{r} + \mathbf{a} \qquad \Rightarrow \qquad \mathbf{r} = \Lambda^{-1} \left(\mathbf{r}' - \mathbf{a} \right).
%
%
\end{align}
Then, by definition, the new field $\phi'(\vx',z')$ evaluated at $\mathbf{r}'$ must take the same value of the original field $\phi(\vx,z)$ evaluated at $\mathbf{r}$, namely 
\begin{align}\label{p190}
\phi'(\br') = \phi(\br) = \phi(\Lambda^{-1}\br' - \Lambda^{-1}\va),
%
%
\end{align}
where we have used the rightmost relation in Eq. \eqref{p180}. Because of the arbitrariness of the point $\br'$, we can drop the prime symbol $(\,'\,)$ and rewrite Eq. \eqref{p190} as
\begin{align}\label{p200}
\phi'(\br) = \phi(\Lambda^{-1}\br - \Lambda^{-1}\va).
%
%
\end{align}
This equation \emph{defines} the behavior of a scalar field under translations and rotations.

Now, suppose that $\phi(\vx, z)$ is a solution of the PWE, namely  
\begin{align}\label{p210}
\left( i \pd{}{z} + \frac{1}{2 k_0}\nabla^2  \right)\phi(\vx,z) = 0.
%
%
\end{align}
 Then the question is: What are the admissible transformations $\left( \va, \Lambda \right)$ such that $\phi(\Lambda^{-1}\br - \Lambda^{-1}\va)$ is still a solution of the PWE? An instructive and elegant method for answering this question without embarking on  calculations of chained partial derivatives, is furnished by the Fourier transform technique. Suppose that the field $\phi(\vx, z)=\phi(\br)$ can be expressed as a \emph{three-dimensional} Fourier integral:
\begin{align}\label{p220}
\phi(\br) = \frac{1}{(2 \pi)^{3/2}} \int (d^3 p ) \, \widetilde{\phi}(p_1,p_2, p_3) e^{i p_\mu x^\mu},
%
%
\end{align}
where  $(d^3 p ) = d p_1 \, d p_2  \, d p_3$. Substituting Eq. \eqref{p220} into Eq. \eqref{p210} we obtain
\begin{align}\label{p230}
 \int (d^3 p ) \, \left(p_3 + \frac{p_1^2 + p_2^2}{2 k_0} \right) \widetilde{\phi}(p_1,p_2, p_3) e^{i p_\mu x^\mu} = 0.
%
%
\end{align}
Thus, the differential equation \eqref{p210} became an algebraic equation in the amplitude $\widetilde{\phi}(p_1,p_2, p_3)$:
\begin{align}\label{p240}
 \left(p_3 + \frac{p_1^2 + p_2^2}{2 k_0} \right) \widetilde{\phi}(p_1,p_2, p_3)  = 0.
%
%
\end{align}
From this equation it follows that the Fourier amplitude $\widetilde{\phi}(p_1,p_2, p_3)$ can be different from zero only when $p_3 + (p_1^2 + p_2^2)/(2 k_0)=0$. This constraint compels $\widetilde{\phi}(p_1,p_2, p_3)$ to have the form
\begin{align}\label{p250}
 \widetilde{\phi}(p_1,p_2, p_3)  = \delta \left(p_3 + \frac{p_1^2 + p_2^2}{2 k_0} \right) \widetilde{\varphi}(p_1,p_2, p_3),
%
%
\end{align}
where, because of the Dirac delta property  $x \, \delta(x)=0$, the  amplitude $\widetilde{\varphi}(p_1,p_2, p_3)$ can be a completely arbitrary smooth function of $(p_1,p_2, p_3)$. 

By definition of Fourier transform and using Eq. \eqref{p220}, we can write
\begin{align}\label{p260}
\phi(\Lambda^{-1}\br - \Lambda^{-1}\va) = & \; \frac{1}{(2 \pi)^{3/2}} \int (d^3 p ) \, 
\widetilde{\phi}(p_1,p_2,p_3)e^{  -i p_\mu (\Lambda^{-1})^\mu_{\phantom{x} \nu}\,a^\nu } e^{ i p_\mu (\Lambda^{-1})^\mu_{\phantom{x} \nu}\,x^\nu }.
%
%
\end{align}
Then, defining the new dummy variable $q_\nu$ as
\begin{align}\label{p270}
q_\nu =  p_\mu (\Lambda^{-1})^\mu_{\phantom{x} \nu} \qquad \Rightarrow \qquad (d^3 p ) \to \abs{\operatorname{det} \Lambda} d q_1 d q_2 dq_3 \equiv \abs{\operatorname{det} \Lambda} (d^3 q )
%
%
\end{align}
permits us to rewrite Eq. \eqref{p260} in the form
\begin{align}\label{p280}
\phi(\Lambda^{-1}\br - \Lambda^{-1}\va) = & \; \frac{1}{(2 \pi)^{3/2}} \int (d^3 q ) \, \left[ \widetilde{\phi}\left( q_\mu\Lambda^\mu_{\phantom{x} 1} , q_\mu\Lambda^\mu_{\phantom{x} 2}, q_\mu\Lambda^\mu_{\phantom{x} 3}\right) \, \abs{\operatorname{det} \Lambda} \,
e^{  -i q_\mu a^\mu } \right] \, e^{ i q_\mu x^\mu },
%
%
\end{align}
where we have inverted the first expression in Eq. \eqref{p270} to write $p_\nu = q_\mu \Lambda^\mu_{\phantom{x} \nu}$. To see whether Eq. \eqref{p280} is a solution of the PWE, we substitute it into Eq. \eqref{p210} to eventually obtain the algebraic equation
\begin{align}\label{p290}
0 = & \;  \left(q_3 + \frac{q_1^2 + q_2^2}{2 k_0} \right) \widetilde{\phi}\left( q_\mu\Lambda^\mu_{\phantom{x} 1} , q_\mu\Lambda^\mu_{\phantom{x} 2}, q_\mu\Lambda^\mu_{\phantom{x} 3}\right) \nonumber \\
= & \;  \left(q_3 + \frac{q_1^2 + q_2^2}{2 k_0} \right)
\delta \left(  q_\mu\Lambda^\mu_{\phantom{x} 3} + \fracd{\left(q_\mu\Lambda^\mu_{\phantom{x} 1}\right)^2 + \left(q_\mu\Lambda^\mu_{\phantom{x} 2}\right)^2}{2 k_0} \right)
 \widetilde{\varphi}\left( q_\mu\Lambda^\mu_{\phantom{x} 1} , q_\mu\Lambda^\mu_{\phantom{x} 2}, q_\mu\Lambda^\mu_{\phantom{x} 3}\right),
%
%
\end{align}
where Eq. \eqref{p250} has been used. Since the displacement vector $\va$  does not enter in Eq. \eqref{p290}, it can take any value. However,  Eq. \eqref{p290} put some limitations on the form of the rotation $\Lambda$, which must evidently satisfy the relation
\begin{align}\label{p300}
q_\mu\Lambda^\mu_{\phantom{x} 3} + \fracd{q_\mu q_\nu \Lambda^\mu_{\phantom{x} i} \Lambda^\nu_{\phantom{x} i}}{2 k_0} = C \left( q_3 + \frac{q_1^2 + q_2^2}{2 k_0} \right) ,
%
%
\end{align}
where $C$ is an irrelevant constant that we arbitrarily fix to $C=1$. Equation \eqref{p300} naturally splits in 
\begin{align}\label{p310}
q_\mu \Lambda^\mu_{\phantom{x} 3} = q_3 \qquad \Rightarrow \qquad \Lambda^3_{\phantom{x} 3} = 1, \quad  \Lambda^i_{\phantom{x} 3} =0,
%
%
\end{align}
and
\begin{align}\label{p320}
q_\mu q_\nu \Lambda^\mu_{\phantom{x} i} \Lambda^\nu_{\phantom{x} i} = q_1^2 + q_2^2 \qquad \Rightarrow \qquad 
\Lambda^3_{\phantom{x} i} =0, \quad \Lambda^k_{\phantom{x} i} \Lambda^l_{\phantom{x} i} = \delta^{kl}.
%
%
\end{align}
The last relation can be simply written as $L L^T = I_2$, where with $L$ we denoted the  $2 \times 2$ \emph{principal submatrix} of $\Lambda$ obtained from the latter deleting the third row an the third column and  $I_2$ is the  $2 \times 2$ identity matrix. The superscript ``$T$'' indicates the transpose of the matrix. 

To summarize, we have found that the transformations that leave the PWE invariant consist of translations by arbitrary three-dimensional vectors $\va$ and of two-dimensional rotations around the $z$-axis of the form
\begin{align}\label{p330}
\Lambda = \begin{pmatrix}
L_{11} & L_{12} & 0 \\
L_{21} & L_{22} & 0 \\
0 & 0 & 1
\end{pmatrix},
%
%
\end{align}
where $L: \, L L^T = I_2$ denotes an arbitrary $2 \times 2$ \emph{orthogonal} matrix.

\subsection{Canonical energy-momentum tensor for the paraxial wave equation}

Given the paraxial Lagrangian 
\begin{align}\label{p340}
\mL =  i \, \delta^{3 \mu} \bphi \partial_\mu \phi  - \frac{1}{2 k_0}  \delta^{\mu \nu}_\text{T} \partial_\mu \bphi \partial_\nu \phi,
%
%
\end{align}
the canonical energy-momentum tensor can be build in the usual manner as
\begin{align}\label{p350}
\mathscr{T}_{\mu \nu} = & \; \pd{\mL}{(\partial^\mu \phi)} \partial_\nu \phi + \pd{\mL}{(\partial^\mu \bphi)} \partial_\nu \bphi  - \delta_{\mu \nu} \mL \nonumber \\
= & \; i \, \delta^{3 \mu} \bphi \partial_\nu \phi - \frac{1}{2 k_0} \delta^{\mu \alpha}_\text{T} 
\bigl( \partial_\alpha \bphi \partial_\nu \phi +   \partial_\alpha \phi \partial_\nu \bphi\bigr)  - \delta_{\mu \nu} \mL.
%
%
\end{align}
Explicitly, we have:
\begin{align}\label{p355}
\mathscr{T} = 
\begin{pmatrix}
  -i \bphi \phi_{\!,z} - \frac{  \abs{\phi_{\!,x}}^2 -   \abs{\phi_{\!,y}}^2}{2 k_0} & - \frac{  \phi_{\!,x}^*  \phi_{\!,y} + \phi_{\!,x}  \phi_{\!,y}^*}{2 k_0} & - \frac{  \phi_{\!,x}^*  \phi_{\!,z} + \phi_{\!,x}  \phi_{\!,z}^*}{2 k_0} \\[0.4em]
 - \frac{  \phi_{\!,x}^*  \phi_{\!,y} + \phi_{\!,x}  \phi_{\!,y}^*}{2 k_0} &   -i \bphi \phi_{\!,z} + \frac{  \abs{\phi_{\!,x}}^2 -   \abs{\phi_{\!,y}}^2}{2 k_0} & - \frac{  \phi_{\!,y}^*  \phi_{\!,z} + \phi_{\!,y}  \phi_{\!,z}^*}{2 k_0} \\[0.4em]
  i \bphi \partial_x \phi & i \bphi \partial_y \phi &    \fracd{1}{2 k_0} \abs{\bnabla \phi}^2
 \end{pmatrix},
%
%
\end{align}
where we have used the shorthand $\phi_{,\mu} = \partial_\mu \phi$.
By definition
\begin{align}\label{p370}
\pd{}{x_\mu}\mathscr{T}_{\mu \nu} = 0, \qquad \text{although} \qquad \pd{}{x_\nu}\mathscr{T}_{\mu \nu} \neq 0.
%
%
\end{align}
This means that there are a conserved energy $H$ and a conserved \emph{transverse} linear momentum $\vP$ defined as
\begin{align}\label{p380}
 H = \int d \vx \, \mathscr{H} \qquad \text{and} \qquad  \vP = \int d \vx \, \mathscr{P}^l \bep_l 
%
%
\end{align}
where
\begin{align}\label{p390}
\mathscr{H} = \mathscr{T}_{33} =\frac{1}{2 k_0} \abs{\bnabla \phi}^2, 
%
%
\end{align}
in agreement with Eq. \eqref{p140}, and
\begin{align}\label{p400}
\mathscr{P}_l = \mathscr{T}_{3 l} = - \bphi\left( \frac{1}{i} \pd{}{x^l} \right) \phi.
%
%
\end{align}
It should be noticed that the minus sign in the equation above, opposite to the sign of $\mathscr{H}$ in Eq. \eqref{p390},  is consistent with the condition implied by the Dirac delta in Eq. \eqref{p250}, because
\begin{align}\label{p405}
 H = & \; \frac{1}{2 k_0}\int d \vx \,  \abs{\bnabla \phi}^2 \nonumber \\
= & \; -\frac{1}{2 k_0}\int d \vx \,  \bphi \nabla^2 \phi \nonumber \\
= & \; -\int d \vx \,  \bphi\left( \frac{1}{i} \pd{}{z} \right) \phi,
%
%
%
\end{align}
where the equation of motion \eqref{p210} has been used. Therefore, one can consider the conserved quantities $( \vP, H) \equiv (P^1,P^2,P^3)$ as the components of a conserved three-momentum $\mathscr{P}^\mu$, where $\mathscr{P}^3 \sim - \mathscr{H}$.

Since $\delta^{ij}_\text{T} = \delta^{ij}$, the transverse part $\mathscr{T}^{ij}$ of $\mathscr{T}^{\mu \nu}$ is symmetric and can be written as
\begin{align}\label{p410}
\mathscr{T}^{ij} = -\frac{1}{2k_0} \Bigl( \partial^i \bphi \partial^j \phi + \partial^i \phi \partial^j \bphi - \delta^{ij}\abs{\bnabla \phi}^2 \Bigr).
%
%
\end{align}
Then, we can construct the conserved tensor density
\begin{align}\label{p420}
\mathscr{M}^{\lambda ij} \equiv  x^i \mathscr{T}^{\lambda j} - x^j \mathscr{T}^{\lambda i},
%
%
\end{align}
such that 
\begin{align}\label{p430}
\partial_\lambda \mathscr{M}^{\lambda ij}  = \mathscr{T}^{ij} -\mathscr{T}^{ji} = 0. 
%
%
\end{align}
Therefore, the quantity 
\begin{align}\label{p440}
J^{ij} \equiv \int d \vx \,  \mathscr{M}^{3 ij},
%
%
\end{align}
is conserved during propagation, namely
\begin{align}\label{p450}
\pd{}{z}J^{ij} =0.
%
%
\end{align}
It is clear that the antisymmetric tensor $J^{ij}$ consists of only one independent parameter, which amounts to the longitudinal component of the orbital angular momentum:
\begin{align}\label{p460}
J_z = & \;  J^{21} \nonumber \\
= & \; \int d \vx \; \phi^*(\vx,z) \left( x \frac{1}{i} \pd{}{y} - y \frac{1}{i} \pd{}{x}\right) \phi(\vx,z).
%
%
\end{align}

\subsection{Internal symmetries}

The Lagrangian Eq. \eqref{p340} is manifestly invariant under the transformation
\begin{align}\label{p461}
\phi \to e^{- i \Lambda} \phi, \qquad \bphi \to e^{i \Lambda} \bphi,
%
%
\end{align}
where $\Lambda$ is a real constant. From the Noether's theorem it follows that there exist a conserved current (see, e.g., Ref. \cite{Greiner}, p. 46, Eq. (2.83))
\begin{align}\label{p462}
\mathscr{J}^\mu = & \; \frac{1}{i} \left[ \phi \pd{\mL}{(\partial_\mu \phi)} - \bphi \pd{\mL}{(\partial_\mu \bphi)} \right] \nonumber \\
= & \; \delta^{\mu 3} \abs{\phi}^2 -\frac{i}{2 k_0} \delta^{\mu \nu}_\text{T} \bigl( \bphi \partial_\nu \phi - \phi \partial_\nu \bphi  \bigr) \nonumber \\
= & \; \delta^{\mu 3} \abs{\phi}^2 -\frac{i}{2 k_0} \delta^{lm} \bigl( \bphi \partial_m \phi - \phi \partial_m \bphi  \bigr)  ,
%
%
\end{align}
where Eq. \eqref{p340} has been used. 
The current $\mathscr{J}^\mu$  has a vanishing three-divergence \cite{Ryder,LowellBrown,Weinberg}
\begin{align}\label{p463}
\partial_\mu \mathscr{J}^\mu = \partial_z \mathscr{J}_z + \bnabla \cdot \pmb{\mathscr{J}}=0,
%
%
\end{align}
namely
\begin{align}\label{p464}
\pd{}{z} \abs{\phi}^2 = \frac{i}{2 k_0}\bnabla \cdot \bigl( \bphi \bnabla \phi - \phi \bnabla \bphi \bigr).
%
%
\end{align}
This continuity equation has the same form, when position $z$ is replaced by time $t$, of the continuity equation for the conservation of probability in the quantum theory of a free two-dimensional particle \cite{Merzbacher}. Moreover, as noticed in \cite{Haus}, Eq. \eqref{p464} is strictly connected to the Poynting theorem in classical electrodynamics \cite{BW}.
Integrating both sides of this equation over all the $xy$-plane we obtain 
\begin{align}\label{p465}
\partial_z \int d \vx \mathscr{J}_z = & \;  - \int d \vx \, \bnabla \cdot \pmb{\mathscr{J}} \nonumber \\
= & \; 0 ,
%
%
\end{align}
where the right side amounts to the two-dimensional integral of a two-divergence
 and then vanishes for fields localized within a \emph{finite} region of the $xy$-plane.
This equation states that during propagation of a monochromatic  optical field along the $z$-axis,  the ``charge'' $Q$ defined as
\begin{align}\label{p466}
Q =  \int d \vx \, \abs{\phi(\vx,z)}^2, \qquad \text{is conserved:} \qquad \pd{Q}{z}=0.
%
%
\end{align}
As we will see later, in the quantum version of the theory this charge simply corresponds to the total number of the particles in the field.

\section{Quantization of the paraxial field}

The quantum theory of electromagnetic fields in the regime of paraxial propagation, has been accomplished by several authors in the past \cite{G&H,D&G,AielloPRA,AielloOL}. In these works, the quantized fields where  \emph{vector fields} obeying Maxwell equations. However, using the full machinery of quantum electrodynamics is not really necessary for many practical applications. Therefore, in the present work we simply aim at quantizing the complex \emph{scalar} field $\phi(\vx,z)$ satisfying the paraxial wave equation \eqref{p110}. In practice, we will follows basically the same procedure outlined in \cite{Schiff,Greiner}, for the quantization of the nonrelativistic Schr\"{o}dinger equation.

We begin by rewriting  the Lagrangian \eqref{p165}
\begin{align}\label{p470}
\mL =  i  \hbar \,\bphi \partial_z \phi  - \frac{\hbar}{2 k_0} \bnabla \bphi \cdot \bnabla \phi ,
%
%
\end{align}
and the \emph{canonically conjugate field} $\Pi(\vx, z)$ associated with $\phi(\vx, z)$:
\begin{align}\label{p480}
\Pi(\vx, z) = \pd{\mL}{(\partial_z \phi)} =i \, \hbar \, \phi^*(\vx, z),
%
%
\end{align}
where the constant multiplicative term $\hbar$ that we added, does not alter the dynamics of the fields and can be eliminated by absorbing it into the definition of the field: $\phi \to \phi/\sqrt{\hbar}$. Moreover, if we multiply both sides of Eq.  \eqref{p470} by the speed of light $c$ and we define the new time-like variable $\tau = z/c$, then the so-obtained Lagrangian 
\begin{align}\label{p482}
\mL \to i  \hbar \, \bphi \pd{\phi}{\tau} - \frac{\hbar \, c}{2 k_0} \bnabla \bphi \cdot \bnabla \phi ,
%
%
\end{align}
becomes identical to the Lagrangian associated to the Schr\"{o}dinger equation of a particle of mass $m = \hbar k_0/c$, whose motion is restricted to the plane $xy$.

Since $\Pi^* = 0$, there are only two independent fields, either $(\phi, \Pi)$ or $(\phi, \bphi)$. We choose the second pair and write the Hamiltonian density \eqref{p390} as
\begin{align}\label{p490}
\mathscr{H}(\vx, z) = \Pi \pd{\phi}{z} - \mL = \frac{\hbar}{2 k_0} \bnabla \bphi \cdot \bnabla \phi .
%
%
\end{align}
As usual, the total Hamiltonian is obtained integrating $\mathscr{H}(\vx, z)$ over the $xy$-plane:
\begin{align}\label{p500}
H = &\; \int d \vx \, \mathscr{H}(\vx, z) \nonumber \\
= &\;   \int d \vx \,  \bphi(\vx, z)\left(  -\frac{\hbar}{2 k_0}\nabla^2 \right) \phi(\vx, z) ,
%
%
\end{align}
where we used integration by part to pass from the first to the second line of Eq. \eqref{p500}.

At this point, the classical theory is quantized by simply promoting the two classical fields $\phi(\vx,z)$ and $\bphi(\vx,z)$, to the operators $\hphi(\vx,z)$ and $\hbphi(\vx,z)$, respectively, and then postulating the \emph{equal-$z$ canonical commutation relations}:
\begin{align}\label{p510}
\Bigl[\hphi(\vx,z),\hbphi(\vx',z) \Bigr] =  \delta(\vx - \vx')
%
%
\end{align}
and
\begin{align}\label{p520}
\Bigl[\hphi(\vx,z),\hphi(\vx',z) \Bigr] = 0 = \Bigl[\hbphi(\vx,z),\hbphi(\vx',z) \Bigr].
%
%
\end{align}
The Hamilton equations of motion now become
\begin{align}\label{p530}
i \hbar \, \pd{}{z} \hphi(\vx,z) = \Bigl[\hphi(\vx,z),\hH \Bigr],
%
\end{align}
where $\hH$ is straightforwardly derived  from Eq. \eqref{p500}:
\begin{align}\label{p540}
\hH =   \int d \vx \,  \hbphi(\vx, z)\left(  -\frac{\hbar}{2 k_0}\nabla^2 \right) \hphi(\vx, z).
%
%
\end{align}
From Eqs. (\ref{p510}-\ref{p540}) it follows that with our choice the Hamiltonian operator $\hH$ must have the dimensions of an energy divided by a velocity, therefore $c \hH$ represents a \emph{true} energy. 
Substituting Eq. \eqref{p540} into Eq. \eqref{p530}, one obtains
\begin{align}\label{p550}
i \hbar \, \pd{}{z} \hphi(\vx,z) =& \;  \Bigl[\hphi(\vx,z),\hH \Bigr] \nonumber \\
=& \;   \int d \vx' \,  \Bigl[\hphi(\vx,z),\hbphi(\vx',z) \Bigr]\left(  -\frac{\hbar}{2 k_0}\nabla^2_{\vx'} \right) \hphi(\vx', z) \nonumber \\
=& \;  -\frac{\hbar}{2 k_0}\nabla^2  \hphi(\vx, z) ,
%
%
\end{align}
which correctly reproduces the PWE.

\subsection{Mode expansion and particle interpretation}

At any position $z$ the fields $\hphi(\vx,z)$ and $\hbphi(\vx,z)$ can be expanded in terms of the \emph{two-dimensional}  Fourier transform representations as 
\begin{align}\label{p560}
\hphi(\vx,z) = \frac{1}{2 \pi} \int d \vp \, \hPhi(\vp,z) e^{i \vp \cdot \vx} 
%
%
\end{align}
and 
\begin{align}\label{p570}
\hbphi(\vx,z) = \frac{1}{2 \pi} \int d \vp \, \hbPhi(\vp,z) e^{-i \vp \cdot \vx}.
%
%
\end{align}
Substituting Eqs. (\ref{p560}-\ref{p570}) into Eq. \eqref{p540} we obtain, after some manipulation,
\begin{align}\label{p580}
\hH = \int d \vp \, \frac{\hbar \, p^2}{2 k_0} \, \hbPhi(\vp,z) \hPhi(\vp,z)  ,
%
%
\end{align}
where $p^2 = \vp \cdot \vp$. Using the Fourier inversion formula, we obtain from Eqs. (\ref{p560}-\ref{p570})
\begin{align}\label{p590}
\hPhi(\vp,z) = \frac{1}{2 \pi} \int d \vx \, \hphi(\vx,z) e^{-i \vp \cdot \vx} 
%
%
\end{align}
and 
\begin{align}\label{p600}
\hbPhi(\vp',z) = \frac{1}{2 \pi} \int d \vx' \, \hbphi(\vx',z) e^{i \vp' \cdot \vx'} .
%
%
\end{align}
Therefore, after a straightforward calculation one finds that the canonical commutation relations (\ref{p510}-\ref{p520}) yield for $\hPhi$ and  $\hbPhi$ the following results:
\begin{align}\label{p610}
\Bigl[\hPhi(\vp,z),\hbPhi(\vp',z) \Bigr] =  \delta(\vp - \vp')
%
%
\end{align}
and
\begin{align}\label{p620}
\Bigl[\hPhi(\vp,z),\hPhi(\vp',z) \Bigr] = 0 = \Bigl[\hbPhi(\vp,z),\hbPhi(\vp',z) \Bigr].
%
%
\end{align}

The $z$-derivative of $\hPhi(\vp,z)$ is given by the Hamilton equation
\begin{align}\label{p630}
i \hbar \, \pd{}{z} \hPhi(\vp,z) =& \;  \Bigl[\hPhi(\vp,z),\hH \Bigr]\nonumber \\
=& \;   \frac{\hbar}{2 k_0}\int d \vq \, q^2 \, \Bigl[\hPhi(\vp,z),\hbPhi(\vq,z)\Bigr] \hPhi(\vq,z)  \nonumber \\
=& \;  -\frac{\hbar \, p^2}{2 k_0} \, \hPhi(\vp,z) ,
%
%
\end{align}
where Eqs. (\ref{p580},\ref{p610}) have been used. This equation can be easily solved to obtain
\begin{align}\label{p640}
 \hPhi(\vp,z) =  \ha(\vp)\,  e^{-i z \, \eta_p  }, \qquad \text{with} \qquad \eta_p \equiv \frac{p^2}{2 k_0},
%
%
\end{align}
where $\ha(\vp) \equiv \hPhi(\vp,0)$. It should be noticed that Eq. \eqref{p640} reproduces  the so-called ``Fresnel-propagation'' law for classical paraxial fields \cite{Goodman}.
A similar calculation also shows that
\begin{align}\label{p645}
 \hbPhi(\vp,z) =  \hac(\vp) \, e^{i z \, \eta_p  },
%
%
\end{align}
with $\hac(\vp) \equiv \hbPhi(\vp,0)$.
 Then, we can rewrite the fields (\ref{p560}-\ref{p570}) as
\begin{align}\label{p650}
\hphi(\vx,z) = \frac{1}{2 \pi} \int d \vp \, \ha(\vp) \, e^{i \vp \cdot \vx - i z \, \eta_p}.
%
%
\end{align}
and
\begin{align}\label{p660}
\hbphi(\vx,z) = \frac{1}{2 \pi} \int d \vp \, \hac(\vp) \, e^{-i \vp \cdot \vx + i z \, \eta_p}.
%
%
\end{align}
From Eq. \eqref{p610} and Eqs. (\ref{p640}-\ref{p645}), it follows that 
\begin{align}\label{p665}
\bigl[\ha(\vp),\hac(\vp') \bigr] =  \delta(\vp - \vp'), \qquad \text{and} \qquad \bigl[\ha(\vp),\ha(\vp') \bigr] = 0 = \bigl[\hac(\vp),\hac(\vp') \bigr].
%
%
\end{align}

\subsubsection{Spectrum of the field}

In the \emph{Fourier representation}, the Hamiltonian \eqref{p580} becomes manifestly $z$-independent:
\begin{align}\label{p670}
\hH = \int d \vp \, \hbar \, \eta_p \, \hac(\vp) \ha(\vp) .
%
%
\end{align}
According to our analysis about the conservation laws associated to the PWE, there must exist a set of three conserved operators $\{ \hP^1, \hP^2, \hP^3\}$, where $\hP^3 \equiv - \hH$ and
\begin{align}\label{p680}
\hP^l = & \; \int d \vx \, \hbphi(\vx,z) \left( \frac{1}{i} \pd{}{x_l} \right) \hphi(\vx,z) \nonumber \\
= & \; \int d \vp \, \hbar \,  p^l \, \hac(\vp) \ha(\vp) ,
%
%
\end{align}
namely:
\begin{align}\label{p685}
\hP^\mu \equiv \{ \hP^1, \hP^2, \hP^3\} =  \int d \vp \, \hbar \,  \left\{ p^1,p^2, p^3=-\frac{p^2}{2 k_0} \right\} \, \hac(\vp) \ha(\vp) .
%
%
\end{align}
The invariance of these operators with respect to $z$-propagation,  can be proved directly by calculating  the commutator
\begin{align}\label{p690}
\bigl[\hP^l ,\hH \bigr]= & \; \int d \vp \int d \vq \, (\hbar \,  p^l) (\hbar \,  \eta_{q})\, 
\bigl[\hac(\vp) \ha(\vp),\hac(\vq) \ha(\vq) \bigr] \nonumber \\
= & \; \int d \vp \int d \vq \, (\hbar \,  p^l) (\hbar \,  \eta_{q})\, \Bigl\{
\hac(\vp)\bigl[\ha(\vp), \hac(\vq) \bigr]\ha(\vq) + \hac(\vq)\bigl[\hac(\vp), \ha(\vq) \bigr]\ha(\vp) \Bigr\}\nonumber \\
= & \; 0,
%
%
\end{align}
where Eq. \eqref{p665} and the commutator distributive law $[AB,CD] = A[B,C] D + A C[B,D] + [A,C] D B + C[A,D] B$, have been used. Proceeding in the same manner, it is not difficult to see that also the ``number'' operator
\begin{align}\label{p700}
\hN =  \int d \vx \, \hbphi(\vx,z) \hphi(\vx,z) = \int d \vp \, \hac(\vp) \ha(\vp)
%
%
\end{align}
is conserved: $\bigl[\hN, \hH \bigr]=0$. Moreover, a straightforward calculations shows that $\bigl[\hP^i, \hP^j]=0$.

From the equations \eqref{p650}, \eqref{p685} and by using the Campbell-Baker-Hausdorff formula, it is not difficult to prove that 
\begin{align}\label{p705}
\hphi(\vx,z)  = e^{- i \hP^\mu x_\mu/\hbar} \, \hphi(\mathbf{0},0) \, e^{i \hP^\nu x_\nu/\hbar}   .
%
%
\end{align}

Since the four operators  $\{ \hP^1, \hP^2, \hH, \hN \}$ commute, they can be simultaneously diagonalized. The procedure to find a complete set of eigenstates of such operators is pretty standard and can be found in many textbooks; therefore now we will only sketch the procedure following Ref. \cite{Lee}. Let $\ket{n'}$ be an \emph{eigenstate} of $\hN$ with \emph{eigenvalue} $n'$:
\begin{align}\label{p710}
\hN \ket{n'} = n' \ket{n'},
%
%
\end{align}
where $n'$ is real number, not necessarily integer. Since 
\begin{align}\label{p720}
\hN \hac(\vp )= & \; \int d \vq \, \hac(\vq) \ha(\vq)\hac(\vp ) \nonumber \\
= & \; \int d \vq \, \hac(\vq) \Bigl( \ha(\vq)\hac(\vp ) - \hac(\vp)\ha(\vq) + \hac(\vp)\ha(\vq) \Bigr) \nonumber \\
= & \; \int d \vq \, \hac(\vq) \bigl[\ha(\vq), \hac(\vp) \bigr] + \int d \vq \,  \hac(\vq)\hac(\vp ) \ha(\vq) \nonumber \\
= & \; \hac(\vp) \bigl( 1 + \hN \bigr) 
%
\end{align}
and 
\begin{align}\label{p730}
\hN \ha(\vp )= & \; \int d \vq \, \hac(\vq) \ha(\vq)\ha(\vp ) \nonumber \\
= & \; \int d \vq \, \Bigl( \hac(\vq) \ha(\vp) - \ha(\vp) \hac(\vq) + \ha(\vp) \hac(\vq) \Bigr)\ha(\vq ) \nonumber \\
= & \;- \int d \vq \, \bigl[\ha(\vp), \hac(\vq) \bigr]\ha(\vq)  + \ha(\vp )\int d \vq \,  \hac(\vq) \ha(\vq) \nonumber \\
= & \; \ha(\vp) \bigl( \hN - 1\bigr), 
%
%
\end{align}
then it follows that
\begin{align}\label{p740}
\hN \hac(\vp ) \ket{n'} = (n'+1)\hac(\vp )\ket{n'} \qquad \text{and} \qquad \hN \ha(\vp ) \ket{n'} = (n'-1)\ha(\vp )\ket{n'} .
%
%
\end{align}
This procedure may be iterated. For example, it is not difficult to see that  
\begin{align}\label{p745}
\hN \hac(\vp ) \hac(\vp' ) = \hac(\vp ) \hac(\vp' )(\hN +2),
%
%
\end{align}
and
\begin{align}\label{p750}
\hN \ha(\vp ) \ha(\vp' ) = \ha(\vp ) \ha(\vp' )(\hN -2),
%
%
\end{align}
which implies that
\begin{align}\label{p760}
 \hN  \hac(\vp ) \hac(\vp' ) \ket{n'} = (n'+2) \hac(\vp ) \hac(\vp' )\ket{n'}
%
%
\end{align}
and 
\begin{align}\label{p770}
 \hN  \ha(\vp ) \ha(\vp' ) \ket{n'} = (n'-2) \ha(\vp ) \ha(\vp' )\ket{n'}.
%
%
\end{align}
After repeating this procedure $n$ times, we find
\begin{align}\label{p780}
 \hN  \hac(\vp_1 ) \hac(\vp_2 )\ldots \hac(\vp_n )  \ket{n'} = (n'+ n) \hac(\vp_1 ) \hac(\vp_2 )\ldots \hac(\vp_n )  \ket{n'}
%
%
\end{align}
and 
\begin{align}\label{p790}
 \hN  \ha(\vp_1 ) \ha(\vp_2 )\ldots \ha(\vp_n )  \ket{n'} = (n'- n) \ha(\vp_1 ) \ha(\vp_2 )\ldots \ha(\vp_n )  \ket{n'}.
%
%
\end{align}
Since, from the definition \eqref{p700} it follows that $\hN$ is an Hermitean operator positive semidefinite, then in Eq. \eqref{p790} we must have $n'- n \geq 0$ for any integer $n$ and any real number $n'$. Therefore, $n'$ must be an \emph{integer} (otherwise the iteration never stops). If in \eqref{p790}  we choose $n'=n$ and  define the \emph{vacuum state} $\ket{0}$ as
\begin{align}\label{p800}
\ket{0} \equiv \ha(\vp_1 ) \ha(\vp_2 )\ldots \ha(\vp_n )  \ket{n},
%
%
\end{align}
then it follows that the vacuum state does not contains particles:
\begin{align}\label{p810}
\hN \ket{0} =0.
%
%
\end{align}
From Eqs. (\ref{p730},\ref{p810}) it follows that
\begin{align}\label{p820}
\hN \ha(\vp ) \ket{0} = & \;  \ha(\vp) \bigl( \hN - 1\bigr)\ket{0} \nonumber \\
 = & \;  - \ha(\vp) \ket{0},
%
%
\end{align}
which is in contradiction with the fact that $\hN$ is positive semidefinite. Therefore, it must be
\begin{align}\label{p830}
\ha(\vp) \ket{0} = 0.
%
%
\end{align}
Finally, putting $n'=0$ in Eq. \eqref{p780}, we obtain
\begin{align}\label{p840}
 \hN \, \hac(\vp_1 ) \hac(\vp_2 )\ldots \hac(\vp_n )  \ket{0} = n \, \hac(\vp_1 ) \hac(\vp_2 )\ldots \hac(\vp_n )  \ket{0},
%
%
\end{align}
which permits us to identify $\ket{\vp_1, \ldots, \vp_n} \equiv \hac(\vp_1 )\ldots \hac(\vp_n )  \ket{0}$ with the state containing $n$ particles. The single-particle state $\ket{\vp} = \hac(\vp)\ket{0}$ is normalized according to
\begin{align}\label{p850}
\brak{\vp}{\vp'} = & \; \bra{0} \ha(\vp) \hac(\vp') \ket{0} \nonumber \\
= & \; \delta(\vp - \vp') \brak{0}{0},
%
%
\end{align}
where we used Eq. \eqref{p665} to rewrite $\ha(\vp) \hac(\vp') = \delta(\vp - \vp')  +  \hac(\vp')\ha(\vp)  $.
From now on, we \emph{assume} that the vacuum state is normalized, that is $\brak{0}{0}=1$. It is not difficult to verify that the two-particle state $\ket{\vp_1,\vp_2} = \hac(\vp_2)\hac(\vp_1)\ket{0}$ has the expected Bosons symmetry with respect to the exchange of particles:
\begin{align}\label{p860}
\brak{\vp_1,\vp_2}{\vp_1',\vp_2'} = & \; \bra{0}\ha(\vp_2)\ha(\vp_1) \hac(\vp_2')\hac(\vp_1') \ket{0} \nonumber \\
= & \; \delta(\vp_1 - \vp_1') \delta(\vp_2 - \vp_2') + \delta(\vp_1 - \vp_2') \delta(\vp_2 - \vp_1'),
%
%
\end{align}
where Eq. \eqref{p665} has been repeatedly used. This calculation can be straightforwardly generalized to the $n$-particle states.

\subsubsection{Physical quantities}

To begin with, we show that the $n$-particle states $\npa$ are actually eigenstates of the physical observables $\{\hP^1, \hP^2, \hH \}$, as previously claimed (for $\hN$ this has been already  shown in Eq. \eqref{p840}). To prove this, first we have to calculate the action of $\ha(\vp)$ upon $\npa$, namely
\begin{align}\label{p870}
\ha(\vp) \npa = & \; \ha(\vp)\hac(\vp_1) \cdots \hac(\vp_n) \ket{0} \nonumber \\
= & \; \delta(\vp - \vp_1) \ket{\vp_2, \vp_3, \ldots, \vp_n} + \delta(\vp - \vp_2) \ket{\vp_1, \vp_3, \ldots, \vp_n} + \ldots  \nonumber \\
& \;  + \delta(\vp - \vp_n) \ket{\vp_1, \vp_2, \ldots, \vp_{n-1}}   \nonumber \\
= & \; \sum_{k=1}^n \delta(\vp - \vp_k) \ket{\vp_1, \ldots,\vp_{k-1},\vp_{k+1}, \ldots, \vp_n} ,
%
%
\end{align}
where Eq. \eqref{p665} has been used $n$ times. Now we are equipped to calculate
\begin{align}\label{p880}
\hH \npa = & \; \int d \vp \, \hbar \, \eta_p \, \hac(\vp) \ha(\vp) \npa \nonumber \\
= & \; \sum_{k=1}^n \hbar \, \eta_{p_k} \, \hac(\vp_k) \ket{\vp_1, \ldots,\vp_{k-1},\vp_{k+1}, \ldots, \vp_n}\nonumber \\
= & \; \left(\sum_{k=1}^n \hbar \, \eta_{p_k} \right) \npa \nonumber \\
\equiv & \; E_n \npa.
%
%
\end{align}
So, $\npa$ is actually an eigenstate of $\hH$ with eigenvalue $E_n$. In a similar manner we obtain
\begin{align}\label{p890}
\hP^l \npa 
= & \; \left(\sum_{k=1}^n \hbar \, p_k^l \right) \npa \nonumber \\
\equiv & \; P_n^l \npa.
%
%
\end{align}

Next, an important quantity to calculate is the so-called propagator for the paraxial wave equation. It is evaluated from the \emph{two-point correlation function} $\langle \hphi(\vx,z) \hbphi(\vx', z') \rangle_0$ defined as 
\begin{align}\label{p900}
\langle \hphi(\vx,z) \hbphi(\vx', z') \rangle_0 = & \; \bra{0}\hphi(\vx,z) \hbphi(\vx', z')\ket{0} \nonumber \\
= & \;  \frac{1}{(2 \pi)^2} \int d \vp \; e^{i \vp \cdot (\vx - \vx')} \, e^{-i(z-z')p^2/(2 k_0)}
\nonumber \\
= & \; \frac{k_0}{2 \pi i}\fracd{e^{i\frac{k_0}{2} \frac{\abs{\vx - \vx'}^2}{z-z'}}}{z-z'},
%
%
\end{align}
which coincides with the so-called Fresnel propagator in paraxial optics \cite{Viccaro}.

As a further step, we introduce the \emph{position states} $\nXt$ defined as
\begin{align}\label{p910}
\nXt = \hbphi(\vx_1,z) \cdots \hbphi(\vx_n,z) \ket{0} \equiv \hat{X}_n(z) \ket{0},
%
%
\end{align}
where $\hat{X}_n(z) \equiv \prod_{k=1}^n \hbphi(\vx_k,z)$.
 Exploiting the fact that $\hH \ket{0} = 0$, it is easy to see that these states obey the Schr\"{o}dinger equation:
\begin{align}\label{p915}
i \hbar \pd{}{z} \nXt = & \; i \hbar \pd{}{z} \hat{X}_n(z)  \ket{0} \nonumber \\
= & \; \bigl[ \hat{X}_n(z), \hH \bigr] \ket{0} \nonumber \\
= & \; - \hH \nXt.
%
%
%
\end{align}

The \emph{wave function} associated with the scalar field is given by the scalar product between position $\ket{\vx;z}$ and momentum $\ket{\vp}$ single-particle states, namely:
\begin{align}\label{p920}
\bra{0} \hphi(\vx,z)  \ket{\vp} = & \; \bra{0} e^{- i \hP^\mu x_\mu/\hbar} \, \hphi(\mathbf{0},0) \, e^{i \hP^\nu x_\nu/\hbar} \ket{\vp} \nonumber \\
 = & \; \bra{0}  \hphi(\mathbf{0},0)  \ket{\vp}\, e^{i p^\nu x_\nu} \nonumber \\
= & \; \frac{1}{2 \pi}\,e^{i \vp \cdot \vx - i z p^2/(2 k_0)}.
%
%
%
\end{align}
If we denote $\ket{\vx} \equiv \ket{\vx;0}$, then Eq. \eqref{p920} shows that we have actually recovered the normalized Fourier basis in a two-dimensional space:
\begin{align}\label{p930}
\brak{\vx}{\vp} =  \frac{1}{2 \pi} \, e^{i \vp \cdot \vx }.
%
%
%
\end{align}

The action of the field operator $\hphi(\vx,z)$ on the position state $\nXt$ is similar to the action of the annihilation operator $\ha(\vp)$ on the momentum state $\npa$, which we have seen in Eq. \eqref{p870}. In the present case we have
\begin{align}\label{p940}
\hphi(\vx,z) \nXt = & \; \hphi(\vx,z) \hbphi(\vx_1,z) \cdots \hbphi(\vx_n,z) \ket{0} \nonumber \\
= & \; \delta(\vx - \vx_1) \ket{\vx_2, \vx_3, \ldots, \vx_n} + \delta(\vx - \vx_2) \ket{\vx_1, \vx_3, \ldots, \vx_n} + \ldots  \nonumber \\
& \;  + \delta(\vx - \vx_n) \ket{\vx_1, \vx_2, \ldots, \vx_{n-1}}   \nonumber \\
= & \; \sum_{k=1}^n \delta(\vx - \vx_k) \ket{\vx_1, \ldots,\vx_{k-1},\vx_{k+1}, \ldots, \vx_n} ,
%
%
\end{align}
where Eq. \eqref{p510} has been used to  write $\hphi(\vx,z)\hbphi(\vy,z) = \hbphi(\vy,z)\hphi(\vx,z) + \delta(\vx - \vy)$.

\subsection{Mode expansion in different bases}

Amongst the solutions of the paraxial wave equations there are the so-called Hermite-Gauss and Laguerre-Gauss modes \cite{SiegmanBook}. Let $u_a(\vx, z)$ be one of such modes, where the shorthand $a$ denotes a multiple index.
By definition,
\begin{align}\label{p960}
\left( i \pd{}{z} + \frac{1}{2 k_0}\nabla^2  \right)u_a(\vx, z) = 0.
%
%
\end{align}
 These modes form a complete and orthonormal basis, namely
\begin{align}\label{p965}
\sum_a  u_a^*(\vx, z) u_a(\vy, z) = \delta(\vx - \vy), \qquad \int d \vx \, u_a^*(\vx, z) u_b(\vx, z) = \delta_{ab}.
%
%
\end{align}
Therefore, we can express the fields $\hphi(\vx,z)$ and $\hbphi(\vx,z)$ in this basis as
\begin{align}\label{p970}
\hphi(\vx,z) = \sum_a \hat{\phi}_a u_a(\vx, z) \qquad \text{and} \qquad \hbphi(\vx,z) = \sum_a \hat{\phi}_a^\dagger u_a^*(\vx, z),
%
%
\end{align}
where
\begin{align}\label{p980}
\hat{\phi}_a = \int d \vx \, u_a^*(\vx, z) \hphi(\vx,z)  \qquad \text{and} \qquad \hat{\phi}_a^\dagger = \int d \vx \, u_a(\vx, z) \hbphi(\vx,z).
%
%
\end{align}
Using Eq. \eqref{p510} it is straightforward to calculate the commutators
\begin{align}\label{p990}
\bigl[ \hat{\phi}_a , \hat{\phi}_b^\dagger \bigr] = \delta_{ab} \qquad \text{and} \qquad \bigl[ \hat{\phi}_a , \hat{\phi}_b \bigr] = 0 = \bigl[ \hat{\phi}_a^\dagger , \hat{\phi}_b^\dagger \bigr].
%
%
\end{align}
From Eq. \eqref{p980} it follows that
\begin{align}\label{p1000}
\hat{\phi}_a \ket{0} = 0 \qquad \text{and} \qquad \hat{\phi}_a^\dagger \ket{0} =  \int d \vx \, u_a(\vx, z) \ket{\vx;z} \equiv \ket{a}.
%
%
\end{align}
The latter relation tells us that the operator $\hat{\phi}_a^\dagger$ creates a particle in the paraxial mode $u_a(\vx, z)$ from the vacuum state. The single-particle states associated to different paraxial modes are automatically orthogonal:
\begin{align}\label{p1010}
\brak{a}{b} = \bra{0} \hat{\phi}_a \hat{\phi}_b^\dagger \ket{0} = \delta_{ab},
%
%
\end{align}
where Eqs. (\ref{p990}-\ref{p1000}) have been used.

From these definitions, any other relation may be straightforwardly calculated.

\subsection{Connection with the physical electromagnetic fields}

The paraxial scalar field $\phi(\vx, z)$ and its corresponding quantum operator $\hphi(\vx, z)$ are not directly physical electromagnetic fields, but must be understood as ``envelope fields'', in the following sense. Consider a
field $\psi(\vx,z)$ obeying the Helmholtz wave equation (HWE)
\begin{align}\label{p1020}
\left( \frac{\partial^2}{ \partial x^2} + \frac{\partial^2}{ \partial y^2} + \frac{\partial^2}{ \partial z^2} + k_0^2 \right)\psi(\vx,z) = 0.
%
%
\end{align}
Such a field can be thought, for example, as one of the three components of either the electric or the magnetic field. 
Without loss of generality, let us define the envelope field  $\phi(\vx,z)$ via the relation
\begin{align}\label{p1030}
\psi(\vx,z) = \phi(\vx,z)e^{i k_0 z}.
%
%
\end{align}
Substituting this expression in the Helmholtz equation \eqref{p1010} yields the \emph{exact} wave equation governing $\phi(\vx,z)$: 
\begin{align}\label{p1040}
\left[\left( \frac{\partial^2}{ \partial x^2} + \frac{\partial^2}{ \partial y^2}  + 2 i k_0 \frac{\partial}{ \partial z}  \right) + \frac{\partial^2}{ \partial z^2} \right]\phi(\vx,z) = 0.
%
%
\end{align}
Now, the \emph{slowly varying envelope approximation} amounts to the assumption 
\begin{align}\label{p1050}
\left| \frac{\partial^2 \phi}{ \partial z^2} \right| \ll k_0 \abs{ \frac{\partial \phi}{ \partial z} }
%
%
\end{align}
and it permits to omit the last term within square bracket in Eq. \eqref{p1040},  which eventually reduces to the paraxial wave equation 
\begin{align}\label{p1060}
\left( \frac{\partial^2}{ \partial x^2} + \frac{\partial^2}{ \partial y^2} + 2 i k_0 \frac{\partial}{ \partial z}  \right)\phi(\vx,z) = 0.
%
%
\end{align}
Therefore, we must think of $\phi(\vx,z)$ as an  approximation of the actual field $\psi(\vx,z) \exp(-i k_0 z)$. Moreover, if we restore the time dependence (for the monochromatic field)  multiplying $\psi(\vx,z)$ by $\exp(- i \omega_0 t)$, with $\omega_0 \equiv c k_0$, then the actual \emph{scalar} field that can be related to a real \emph{vector} electromagnetic field, is given by
\begin{align}\label{p1070}
\psi(\vx,z;t) = \phi(\vx,z) e^{i k_0 z - \omega_0  t}.
%
%
\end{align}

From this scalar field, the actual electric and magnetic fields may be calculated within the paraxial approximation, as
\begin{align}\label{p1080}
\vec{E} = \frac{iA_0 }{k_0}\left[ \vn +\frac{i}{k_0} \, \bep_3 \,\left( \vn \cdot \bnabla \right) \right]\psi(\vx,z;t) + \text{c.c.},
%
%
\end{align}
and 
\begin{align}\label{p1090}
\vec{B} = \frac{i A_0 }{\omega_0} \left[ \bep_3 \times \vn  + \frac{i}{k_0} \, \bep_3 \,\left( \vn \times \bnabla \right)_3 \right]\psi(\vx,z;t) + \text{c.c.},
%
%
\end{align}
where ``c.c.'' stands for ``complex conjugate'' and $\vn = n_1 \bep_1 + n_2 \bep_2$ is a two-dimensional transverse unit vector that fixes the polarization of the field \cite{D&G}. $A_0$ is a constant real amplitude with the physical dimensions of an electric field.

For the quantum fields, the generalization of the equations above is straightforward. For example, the electric field operator will be written as $\hat{\vec{E}} = \hat{\vec{E}}^{+} + \hat{\vec{E}}^{-}$, where $\hat{\vec{E}}^{-}$ is the Hermitean conjugate of $\hat{\vec{E}}^{+}$. Then, using Eq. \eqref{p650} we can readily write
\begin{align}\label{p1100}
\hat{\vec{E}}^{+} = & \;\frac{i A_0 }{k_0} \left[ \vn +\frac{i}{k_0} \, \bep_3 \,\left( \vn \cdot \bnabla \right) \right]\hphi(\vx,z) e^{i k_0 z - \omega_0  t} \nonumber \\
= & \;  \frac{i  A_0 }{2 \pi k_0} \, e^{i k_0 z - \omega_0  t}\int d \vp \, \ha(\vp) \left( \vn +\bep_3 \,\frac{i}{k_0} \, \vn \cdot \bnabla  \right) \exp \left(i \vp \cdot \vx  -i z \frac{p^2}{2 k_0}\right).
%
%
%
\end{align}
This expression is in full agreement  with equations (17) and (18) of Ref. \cite{Aiello10}, when the latter are reduced to the monochromatic case.

\section{Conclusions}

In this work we have studied the paraxial wave equation from a field-theoretic point of view. We began writing a  Lagrangian apt to yields the Euler-Lagrange equations correctly reproducing the  paraxial wave equation. Then, we studied the symmetries of the latter  and we deduced several conservation laws. Then, we quantized the fields and calculated the relevant physical observables. Finally, we compared our results with previously established ones finding full agreement.

%
%

%
\end{document}